\documentclass[preprintnumbers,article,amsmath,amssymb,floatfix,9pt,prd,twocolumn,superscriptaddress,nofootinbib,amsfonts,showpacs]{revtex4}
\usepackage{amsmath}
\usepackage{amssymb}
\usepackage{graphicx}
\usepackage{subfigure,accents}
\usepackage{booktabs}
\usepackage{rotating}
\usepackage{longtable}
\usepackage{bm}
\usepackage[pagebackref=false,colorlinks=,linkcolor=blue,citecolor=blue]{hyperref}
\usepackage{epstopdf}
 \pdfoutput=1
 \usepackage[utf8]{inputenc}
\usepackage{cases}
\usepackage{amsmath,float}
\usepackage{amssymb}
\usepackage{amsfonts}
\usepackage{amssymb}
\usepackage{dcolumn}
\usepackage{bm}
\usepackage{bbm}
\usepackage{graphicx}
\usepackage{xcolor}
\usepackage{array}
\usepackage{subfigure}
\usepackage{hyperref}
\usepackage{wasysym}

\newcommand{\be}{\begin{equation}}
\newcommand{\ee}{\end{equation}}
\newcommand{\ba}{\begin{eqnarray}}
\newcommand{\ea}{\end{eqnarray}}

\newcommand{\gsim}{\mathrel{\hbox{\rlap{\lower.55ex \hbox {$\sim$}}
                   \kern-.3em \raise.4ex \hbox{$>$}}}}
\newcommand{\lsim}{\mathrel{\hbox{\rlap{\lower.55ex \hbox {$\sim$}}
                   \kern-.3em \raise.4ex \hbox{$<$}}}}

\hypersetup{colorlinks=true,
            breaklinks=true,
            pdfstartview=Fit,
            linkcolor=blue,
            citecolor=blue,
            urlcolor=blue}
            
 \usepackage{graphicx}

\usepackage{amsmath}

\usepackage{braket}

\newcommand{\bw}{\begin{widetext}}
\newcommand{\ew}{\end{widetext}}

\usepackage{mathtools}

\def\ber{\begin{eqnarray}}
\def\eer{\end{eqnarray}}
\def\beq{\begin{equation}}
\def\eeq{\end{equation}}

\usepackage{orcidlink}

\begin{document}

\title{Possible wormholes in generalized geometry-matter coupling gravity induced by the Dekel-Zhao dark matter profile}

\author{A. Errehymy\orcidlink{0000-0002-0253-3578}}
\email{abdelghani.errehymy@gmail.com}
\affiliation{Astrophysics Research Centre, School of Mathematics, Statistics and Computer Science, University of KwaZulu-Natal, Private Bag X54001, Durban 4000, South Africa}
\affiliation{Center for Theoretical Physics, Khazar University, 41 Mehseti Str., Baku, AZ1096, Azerbaijan}

\author{O. Donmez}
\email{orhan.donmez@aum.edu.kw}
\affiliation{College of Engineering and Technology, American University of the Middle East, Egaila 54200, Kuwait}

\author{A. Syzdykova}
\email{syzdykova$_$am@mail.ru}

\author{K. Myrzakulov}
\email{krmyrzakulov@gmail.com}
\affiliation{Department of General and Theoretical Physics, L.N. Gumilyov Eurasian National University, Astana 010008, Kazakhstan }

\author{S. Muminov}
\email{sokhibjan.muminov@mamunedu.uz}
\affiliation{Mamun University, Bolkhovuz Street 2, Khiva 220900, Uzbekistan}

\author{A. Dauletov}
\email{a.dauletov@afu.uz}
\affiliation{
Alfraganus University, Yukori Karakamish Street 2a, Tashkent 100190, Uzbekistan}

\author{J. Rayimbaev}
\email{javlon@astrin.uz}
\affiliation{New Uzbekistan University, Movarounnahr Street 1, Tashkent 100007, Uzbekistan}
\affiliation{Urgench State University, Kh. Alimjan Str. 14, Urgench 221100, Uzbekistan}

\begin{abstract}
In the late 1980s, Morris and Thorne led in theoretical physics by creating solutions to wormholes and formulating the crucial requirements for safe traversability of wormholes. They found that exotic matter must meet the requirement $P_r + \rho < 0$, where $P_r$ is radial pressure and $\rho$ is energy density. This is a rudimentary grasp of our understanding of general relativity. In this paper, we continue their excellent work by looking at how to build traversable wormhole solutions in an extended theory of gravity. We adopt a process of linearly modifying the matter Lagrangian and the energy-momentum tensor with some coupling strengths $\lambda$ and $\chi$. This may be considered as a special case of linear $f(R, T)$ gravity with matter coupling variability or as an additively separable simple $f(R, L_m, T)$ model. We undertake a detailed analysis of static wormhole solutions with a constant redshift function. This allows us to present our results as a first-order approximation in the $f(R, L_m, T)$ scenario. We derive the wormhole shape function from the Dekel-Zhao dark matter distribution in such a way that our solutions satisfy the needed conditions for traversability as well as the requirement of exotic matter. This is particularly exciting as it shows that wormholes in $f(R, L_m, T)$ gravity can sustain both exotic as well as ordinary matter. To ensure that the shape function meets the requirement of flaring-out and is asymptotically flat, we place some constraints on the couplings. We also examine the gravitational lensing effects, which exhibit a repulsive gravitational force that appears in our extended gravity for positive couplings. Lastly, we verify the stability of our wormhole solutions in the Tolman-Oppenheimer-Volkoff (TOV) formalism, solidifying their theoretical support and opening future research doors ajar for further exploration into the fascinating realm of gravitational physics.

\end{abstract}

\maketitle

\section{Introduction} \label{sec:int}
The $\Lambda$CDM model, which is our go-to framework for understanding the universe, has some significant issues that can’t be ignored. Ironically, the cosmological constant, $\Lambda$—critical to this model—also highlights its biggest problems. While it claims to explain the accelerating expansion of the universe \cite{Riess:1998,Perlmutter:1999}, it faces a major hurdle known as ``the worst theoretical prediction in the history of physics'' \cite{Hobson:2006}: the cosmological constant problem \cite{Weinberg:1989}. This discrepancy between what theory suggests and what we actually observe raises real doubts about the model’s reliability and prompts us to rethink our understanding of the cosmos. However, cold dark matter (CDM) is a fascinating yet puzzling aspect of our universe. This strange form of matter doesn't interact with electromagnetic forces, making it completely invisible and undetectable with our current tools. Despite this, it's crucial for understanding how galaxies and galaxy clusters form and evolve, helping to make sense of the observations we see \cite{Salucci:2019, Limousin:2022, Montes:2019}. Even after many expensive attempts, scientists have yet to find any particles that could be associated with dark matter \cite{Smith:1990, Boehm:2004, Mayet:2016}. Recent data from the Planck Satellite \cite{Planck:2020} reveals that about 95\% of the universe is made up of dark energy and dark matter. This reality, along with the unresolved questions surrounding the cosmological constant--which acts as the ``dark energy'' in the $\Lambda$CDM model--raises important doubts about how well this model can truly explain the universe we live in.

The $\Lambda$CDM model faces some tough challenges that go beyond what we've already discussed. Two major issues are the missing satellites problem \cite{Klypin:1999, Moore:1999} and the core/cusp problem \cite{Dubinski:1991, Walker:2011}. The missing satellites problem reveals a significant gap: the model suggests there should be many more halo substructures than we actually observe in the form of satellite galaxies. Meanwhile, the core/cusp problem points out a key inconsistency. While the model predicts that halos should have a sharply peaked (cuspy) density profile, observations of dwarf and low surface brightness galaxies show that their profiles are much flatter (cored). For a deeper look into these important issues and other fundamental challenges with the $\Lambda$CDM model, we recommend checking out \cite{Bull:2016}.

To address the challenges we've discussed, it's worth looking into extended gravity theories as a promising alternative. The main reason for exploring extended gravity theories is their potential to enhance our understanding of the dark sector of the universe by introducing extra gravitational terms that aren't part of Einstein's general theory of relativity (GR), which supports the $\Lambda$CDM model. In what follows, we'll take a closer look at some interesting examples of extended gravity theories and what they can offer.

One of the most exciting extended gravity theories making waves today is the $f(R)$ gravity theory, where $R$ represents the Ricci scalar. For those curious, check out \cite{DeFelice:2010}, which offers a thorough review of the subject. The core idea of $f(R)$ gravity is pretty fascinating: it takes the Einstein-Hilbert action
\begin{equation}\label{eq1}
  S = \frac{1}{16\pi} \int R \sqrt{-g} \, d^4x,   
\end{equation}
where $g$ is the metric determinant, and replaces $R$ with a function of $R$, known as $f(R)$. This shift allows us to rethink gravitational interactions (using units where the speed of light $c$ and Newton's gravitational constant $G_N$ are both set to 1). As a result, the field equations that emerge from this theory include extra terms compared to the familiar equations of GR:
\begin{equation}
G_{\mu\nu} = 8\pi T_{\mu\nu}, 
\end{equation}
where $G_{\mu\nu}$ is the Einstein tensor and $T_{\mu\nu}$ represents the energy-momentum tensor. These added terms might offer a new perspective on dark energy and dark matter—not as mysterious fluids filling the universe, but as crucial modifications to how we understand gravity \cite{Amendola:2007, Nojiri:2006, Capozziello:2005, Capozziello:2012, Boehm:2008}.  However, it's important to note that $f(R)$ gravity has its share of challenges. Many flaws and limitations have already been highlighted in a recent study \cite{Casado:2023}. However, it's crucial to recognize that $f(R)$ gravity faces its own challenges, as highlighted in a recent study \cite{Casado:2023}; nevertheless, extensive investigations have been undertaken on astrophysical configurations \cite{Aman:2025iye, Khan:2024fuh, Khan:2024vsh, Albalahi:2024ujg, Yousaf:2024fkr, Yousaf:2025fkr, Yousaf:2024src, Yousaf:2025svv, Bhatti:2021oyn}, including self-gravitating fuzzy droplets and fuzzy black holes with spherically symmetric static structures, along with several captivating studies on wormholes \cite{Lobo:2009ip, Bahamonde:2016ixz, Karakasis:2021tqx}.

In the realm of alternative theories of gravity that extend GR, the topic of energy conditions becomes particularly nuanced. The additional degrees of freedom in these extended theories of gravity can be interpreted as generalized \textit{effective fluids}, which differ fundamentally from the standard matter fluids typically used as sources in field equations \cite{Capozziello:2011kj}. This perspective has been thoroughly examined in the literature, where energy conditions have been instrumental in constraining various models, including $f(R)$ theories of gravity \cite{PerezBergliaffa:2006ni, Santos:2007bs}, extensions with nonminimal curvature-matter couplings \cite{Bertolami:2009cd, Garcia:2010xb}, and modified Gauss-Bonnet gravity \cite{Garcia:2010xz}, especially with nonminimal matter couplings. Additionally, the proposed $f(R, T)$ gravity models, where $T$ represents the trace of the energy-momentum tensor and $R$ is the curvature scalar, have also been scrutinized using energy conditions \cite{Harko:2011kv}. Constraints have been established for the $f(R, T, R_{\mu\nu}T^{\mu\nu})$ extension \cite{Odintsov:2013iba}, modified teleparallel gravity \cite{Liu:2012fk}, and violations of the null energy condition in bimetric gravity \cite{Baccetti:2012re}. While standard fluids, like perfect matter fluids, generally adhere to well-defined equations of state, allowing us to define thermodynamic quantities such as the adiabatic index and temperature, these \textit{fictitious} fluids may be linked to scalar fields or other gravitational degrees of freedom. This can lead to physical properties that are poorly defined, causing the energy conditions to behave unpredictably compared to their behavior in GR. The implications of this situation can be quite serious. It may result in significant issues with the causal and geodesic structures of the theory, and the energy-momentum tensor could fail to align with the Bianchi identities and conservation laws.

Recently, the authors \cite{Haghani:2021} introduced a new theory called $f(R, L_m, T)$ gravity, which replaces the traditional curvature scalar $R$ with a function $f(R, L_m, T)$ in (\ref{eq1}), providing fresh insights into gravity. In this study, we explore the construction of traversable wormhole solutions within this framework, influenced by the DZ dark matter profile. These structures, which allow rapid travel between stars \cite{Morris:1988}, must lack horizons, and the matter shaping them challenges conventional energy conditions in GR. Although we haven't yet observed wormholes in reality, researchers have come up with some intriguing ideas for how we might detect them. For example, theories like braneworld models \cite{Curiel:2017}, Born-Infeld gravity \cite{Rosa:2022}, Einstein-Cartan theory \cite{Carr:2010}, and $f(R,T)$ theory \cite{Bronnikov:2020} suggest that traversable wormholes could be supported by regular matter—matter that meets the usual energy conditions. One intriguing possibility for the existence of wormholes comes from exploring extended or alternative theories of gravity \cite{Capozziello:2011kj, Errehymy:2023rsm}. These theories offer new perspectives that might help us understand how such fascinating structures could exist in our universe. Nandi and colleagues \cite{Nandi:2017} proposed using gravitational wave ring-down waveforms and gravitational lensing to distinguish between wormholes and black holes. Additionally, Ohgami and Sakai \cite{Ohgami:2015} suggested a method for detecting wormholes surrounded by non-obscuring dust, while Paul and his co-workers \cite{Paul:2020} conducted simulations of thin accretion disks around rotating wormholes, revealing intriguing differences from black hole images.

This paper is organized into several sections for clarity. In Sec. \ref{ch: II}, we introduce the framework of $f(R, L_m, T)$ gravity. Then, in Sec. \ref{ch: III}, we discuss how energy conditions apply to our study. Sec. \ref{ch: IV} is dedicated to exploring the exact solutions for wormholes surrounded by DZ profile halos. To highlight the physical implications of our findings, Sec. \ref{ch: V} looks at the effects of gravitational lensing. In Sec. \ref{ch: VI}, we conduct a stability analysis based on the Tolman-Oppenheimer-Volkoff (TOV) equation. Finally, Sec. \ref{ch: VII} wraps things up with a summary of our conclusions and key insights.

\section{The $f(R,L_m, T)$ gravity}\label{ch: II}

Recent research into compact objects has brought attention to the fascinating theory of $f(R, L_m, T)$ gravity, introduced by the authors \cite{Haghani:2021}. This theory stands out because it expands and brings together existing gravitational models, specifically $f(R, T)$ and $f(R, L_m)$. To clarify the terms, $R$ refers to the Ricci scalar, $T$ is the trace of the energy-momentum tensor $T_{\mu\nu}$, and $L_m$ represents the matter Lagrangian. In the framework of $f(R, L_m, T)$ gravity, the gravitational Lagrangian is expressed as a function that encompasses these variables, written as $L_{grav} = f(R, L_m, T)$. When we look at the complete action for this theory, we can simplify things by assuming natural units where $G_N = c = 1$. Here's how it's formulated \cite{Haghani:2021}:
\begin{equation}\label{frl1}
    S=\frac{1}{16\pi}\int d^4xf(R,L_m, T)\sqrt{-g}+\int d^4xL_m \sqrt{-g}.
\end{equation}
Here, we start by defining $g$ as the determinant of the metric tensor $g_{\mu\nu}$. The stress-energy tensor, which plays a crucial role in our study, can be expressed with the following formula:
\begin{equation}
    T_{\mu \nu}=-\frac{2}{\sqrt{-g}}\frac{\delta(\sqrt{-g}L_m)}{\delta g^{\mu \nu}}.
\end{equation}
Next, we want to understand how the trace $T = T^{\mu}_{\mu}$ of the stress-energy tensor varies with respect to the metric tensor. We can express this relationship as:
\begin{eqnarray}
    \frac{\delta (g^{\alpha \beta}T_{\alpha \beta})}{\delta g^{\mu \nu}}=T_{\mu \nu}+\Theta_{\mu \nu}
\end{eqnarray}
where $\Theta_{\mu \nu}$ is defined as:
\begin{eqnarray}\label{theta}
    \Theta_{\mu \nu}\equiv g^{\alpha \beta}\frac{\delta T_{\alpha \beta}}{\delta g^{\mu \nu}}=L_mg_{\mu \nu}-2T_{\mu \nu}-2g^{\alpha \beta}\frac{\partial^{2}L_m}{\partial g^{\mu \nu} \partial g^{\alpha \beta}}
\end{eqnarray} 
In this paper, we assume the matter Lagrangian takes the form:
\begin{eqnarray}
    L_m=-\rho
\end{eqnarray}
where $\rho$ represents the energy density. We further assume that this energy density $\rho$ depends only on the metric tensor $g_{\mu\nu}$ and not on its derivatives. This assumption helps us focus on the essential geometric features of spacetime described by $g_{\mu\nu}$. As a result, the term $2g^{\alpha \beta} \frac{\partial^{2} L_m}{\partial g^{\mu \nu} \partial g^{\alpha \beta}}$ drops out. With these simplifications in mind, we can rewrite Eq. \eqref{theta} as:
\begin{equation}
    \Theta_{\mu \nu}=-2T_{\mu \nu}-\rho g_{\mu \nu}
\end{equation}
In $f(R, L_m, T)$ gravity, we can derive the field equations by varying the action with respect to the inverse metric $g^{\mu\nu}$. This leads us to the following set of equations:
\begin{eqnarray}\label{frl2}
  &&8\pi T_{\mu\nu}+f_T\tau_{\mu\nu}\nonumber+\frac{1}{2}\left[f_{L_m}+2f_T\right]\left[T_{\mu\nu}-L_mg_{\mu\nu}\right]=-\frac{1}{2}fg_{\mu\nu} \nonumber\\
    &&+\left[R_{\mu\nu}+g_{\mu\nu}\Box-\nabla_\mu\nabla_\nu\right]f_R
\end{eqnarray}
Here, $R_{\mu\nu}$ is the Ricci tensor, and we define the various derivatives as follows:
\begin{eqnarray*}
f_R&\equiv& \frac{\partial f}{\partial R},~~
f_{L_m}\equiv \frac{\partial f}{\partial 
{L_m}},~~
f_{T}\equiv \frac{\partial f}{\partial T},\\ 
\tau_{\mu\nu}&\equiv&2g^{\alpha\beta}\frac{\partial^2L_m}{\partial g^{\mu\nu}\partial g^{\alpha\beta}}.
\end{eqnarray*}

The specific form of the function $f(R, L_m, T)$ significantly influences the gravitational dynamics. For instance:
\begin{itemize}
    \item If we set $f(R, L_m, T) = f(R)$, the equations simplify to those of $f(R)$ gravity.
    \item Choosing $f(R, L_m, T) = f(R, T)$ gives us the field equations for $f(R, T)$ gravity.
    \item If we take $f(R, L_m, T) = f(R, L_m)$, we obtain the $f(R, L_m)$ theory.
    \item Setting $f(R, L_m, T) = R$ leads us back to the standard Einstein field equations:
\end{itemize}
\begin{eqnarray}
    8\pi T_{\mu\nu} &=& R_{\mu\nu}-\frac{1}{2}g_{\mu\nu}R 
\end{eqnarray}

From Eq. (\ref{frl2}), we can see that taking its covariant derivative is likely to produce a non-null result for the covariant derivative of the energy-momentum tensor. This characteristic is commonly found in gravity theories that allow for a coupling between geometry and matter, such as the previously mentioned $f(R, L_m)$ and $f(R, T)$ theories. Essentially, this non-conservation of the energy-momentum tensor would indicate matter creation during the universe's evolution, a topic that has been explored extensively in these theories \cite{Harko:2014pqa}. It's also worth noting a mechanism that can avoid energy-momentum nonconservation, as discussed in \cite{dosSantos:2018nmu}. In that study, the authors examined the equation for $\nabla_\mu T^{\mu\nu}$ within the $f(R, T)$ framework, imposed the condition $\nabla_\mu T^{\mu\nu} = 0$, and derived the corresponding solution for the $f(R, T)$ functional. A similar approach could be applied to the current $f(R, L_m, T)$ gravity in future research.

Now, if we consider an additive structure for the function, such as $f(R, L_m, T) = R + \lambda L_m + \chi T$, where $L_m = -\rho$ (the matter Lagrangian), we can rewrite the field equations as
\begin{equation}\label{frl3}
G_{\mu\nu}=\left[8\pi+\frac{\lambda}{2}+\chi\right] T_{\mu \nu}+\frac{\chi}{2}\left[2\rho+T\right]g_{\mu\nu}.  
\end{equation}
We can also express these equations in a more compact form:
 \begin{equation}
 G_{\mu \nu}=R_{\mu \nu}-\frac{1}{2}Rg_{\mu \nu}=\tilde{T}_{\mu \nu},
 \end{equation}
 where $\tilde{T}{\mu\nu} = T^{(m)}{\mu\nu} + \hat{T}^{(m)}{\mu\nu}$ defines the effective stress-energy tensor. The additional term $\hat{T}^{(m)}{\mu\nu}$ is given by:
 \begin{equation}
 \hat{T}^{(m)}_{\mu \nu}=\left[\frac{\lambda}{2}+\chi\right]T_{\mu \nu}+\frac{\chi}{2}\left[ -2 L_m + T\right]g_{\mu \nu}    
 \end{equation}
Now, when we consider the matter content of a wormhole, we can incorporate an anisotropic fluid source. This is characterized by a stress-energy tensor that satisfies certain energy conditions, given by:
\begin{equation}\label{EMT}
    T_{\mu \nu}=(\rho+P_{t})u_{\mu}u_{\nu}+P_{t}g_{\mu \nu}+(P_{r}-P_{t})v_{\mu}v_{\nu}.
\end{equation}
where $u_\mu$ is a time-like vector and $v_\mu$ is a space-like vector that is orthogonal to $u_\mu$. Under these assumptions, we can express the stress-energy tensor in a diagonal form:
\begin{equation}\label{cw8}
T^{\mu}_{\nu}=\text{diag}\left[-\rho, P_{r}, P_{t}, P_{t}\right],
\end{equation}
where $\rho$ represents the matter-energy density of the wormhole, $P_r$ is the radial pressure, and $P_t$ is the tangential pressure. From our earlier assumptions about $f(R, L_m, T)$, we can derive:
\begin{eqnarray}
T&=&-\rho +P_{r}+2P_{t},
\end{eqnarray}
and consequently, the matter stress-energy tensor can be expressed as:
\begin{eqnarray}
8\pi T_{\mu \nu}&=&T^{(m)}_{\mu \nu}.
\end{eqnarray}

\section{Energy Conditions}\label{ch: III}
To ensure that solutions to Einstein's equations make sense physically, we need to apply energy conditions (ECs). These conditions are crucial for validating the structure of spacetime. The four main ECs---the null energy condition (NEC), weak energy condition (WEC), dominant energy condition (DEC), and strong energy condition (SEC)---help confirm that these solutions align with our understanding of physical laws.

When it comes to the WEC, violations are particularly interesting because they suggest the presence of exotic matter, especially in the context of wormholes (WHs). This violation can be expressed with the formula
\begin{eqnarray}
    T_{\xi \varrho}u^{\xi} u^{\varrho}\geq 0,
\end{eqnarray}

where $T_{\xi \varrho}$ is the energy-momentum tensor, and $u^{\xi}$ is any timelike vector. Researchers Hochberg and Visser \cite{Hochberg:1998ha, Hochberg:1998ii}, building on earlier work by Morris and Thorne \cite{Morris:1988cz}, found that the throat region of a wormhole does violate the NEC. The behavior of expansion $\Omega$, rotation $w_{\xi \varrho}$, and shear $\sigma_{\xi \varrho}$ associated with a vector field $v^{\xi}$ is described by the Raychaudhuri equations, which can be written as:
\begin{eqnarray}
\frac{d \Omega}{d \tau} &=& - \frac{\Omega^2}{3}-R_{\xi \varrho} v^{\xi} v^{\varrho}-\sigma_{\xi \varrho} \sigma^{\xi \varrho}-w_{\xi \varrho} w^{\xi \varrho},\label{aa}\\
\frac{d \Omega}{d \tau} &=&- \frac{\Omega^2}{2}-R_{\xi \varrho} u^{\xi} u^{\varrho}-\sigma_{\xi \varrho} \sigma^{\xi \varrho}-w_{\xi \varrho} w^{\xi \varrho}.
\end{eqnarray}
For gravity to pull objects together, a specific condition must be met: the expression 
\begin{eqnarray}
R_{\xi \varrho} v^{\xi} v^{\varrho} \geq 0
\end{eqnarray}
should hold true for all hypersurface orthogonal congruences. In simpler terms, this means that when certain conditions are satisfied—where \(w_{\xi \varrho} \equiv 0\) and \(\sigma^2 = \sigma_{\xi \varrho} \sigma^{\xi \varrho}\)—we can express this requirement as 
\begin{eqnarray}
T_{\xi \varrho} v^{\xi} v^{\varrho} \geq 0.
\end{eqnarray}

The energy conditions (ECs), which include the null energy condition (NEC), weak energy condition (WEC), dominant energy condition (DEC), and strong energy condition (SEC), are derived from Einstein's field equations. These conditions set specific limits on the components of the energy-momentum tensor, which relate to how energy and momentum are distributed. They can be described as follows:

\begin{enumerate}
    \item \textbf{NEC}: The sum of pressure \(p_{i}\) and energy density \(\rho\) must be greater than or equal to zero:
   \begin{eqnarray}\label{NEC}
   (p_{i} + \rho) \geq 0.
   \end{eqnarray}
    \item  \textbf{WEC}: The energy density \(\rho\) must be non-negative, and the sum of pressure and energy density must also be non-negative:
    \begin{eqnarray}
   \rho \geq 0, \quad (p_{i} + \rho) \geq 0.
   \end{eqnarray}
    \item \textbf{DEC}: The difference between energy density and the absolute value of pressure must be non-negative:
    \begin{eqnarray}
   (\rho - |p_{i}|) \geq 0,
   \end{eqnarray}
    \item \textbf{SEC}: The sum of energy density and the total pressure must be non-negative, along with the previous condition:
    \begin{eqnarray}\label{SEC}
   (\rho + \sum_{i=1}^{3} p_{i}) \geq 0, \quad (p_{i} + \rho) \geq 0.
   \end{eqnarray}
    where \(i\) refers to spatial indices.
\end{enumerate}

When we encounter violations of energy conditions, it often points to the existence of exotic matter. For instance, wormhole configurations specifically require a breach of the NEC, which is the least strict among the various energy conditions. Studies have shown that the stability of static wormholes relies on the presence of this exotic matter. However, it's essential to highlight that no experimental evidence has yet confirmed the existence of such matter, which raises questions about the actual feasibility of wormholes in the physical world. In the field of modified gravity, Cappozziello et al. \cite{Capozziello:2013vna} have delved deeply into the implications of these energy conditions. This topic is quite intricate, as many researchers interpret the gravitational field equations as effective Einstein equations. In these theories, the additional degrees of freedom can be viewed as generalized geometrical fluids, which are quite different from the standard matter fluids we typically see in field equations \cite{Capozziello:2011kj}. While ordinary fluids usually follow clear equations of state, these so-called ``fictitious'' fluids might be linked to scalar fields or other gravitational elements, particularly in contexts like $f(R)$ gravity. This can lead to some confusing scenarios regarding their physical properties, potentially causing serious issues. Such ambiguities might create gaps in the causal and geodesic structures of the theory and could lead to inconsistencies between the energy-momentum tensor and the fundamental conservation laws. Because of this, we felt it necessary to issue a cautionary note regarding findings in the existing literature \cite{ Santos:2007kj, Bertolami:2009kj, Akramov:2024bqp, Rakhmanov:2025bqp, Turimov:2025tmf, Razina:2011wv, Kumar:2023cmo}. While we can express the weak, null, dominant, and strong energy conditions similarly to how they are framed in GR, their meanings can change significantly due to possible alterations in causal and geodesic structures and gravitational interactions. In this light, we intend to use the energy conditions to set constraints on a newly proposed gravity theory known as $f(R, L_m, T)$ \cite{Haghani:2021}. These energy conditions place important restrictions on how matter and energy are distributed in spacetime, ensuring that the energy-momentum tensor remains positive and that gravity retains its attractive nature. The significance of these conditions has been recognized in the literature for quite some time, as seen in works like \cite{Tipler:1978zz, Morris:1988tu}. Some references, such as \cite{Wang:2012rw}, specifically explore how energy conditions apply in extended gravity theories. Additionally, it’s worth noting other important findings regarding energy conditions in alternative gravity theories. For example, \cite{Atazadeh:2009kj} looks into energy conditions in $f(R)$ gravity and Brans-Dicke theories, while \cite{Albareti:2009kj} examines the non-attractive nature of gravity in $f(R)$ models. Furthermore, studies of energy conditions in the Jordan frame are discussed in \cite{Chatterjee:2013kj}.
\section{Exact solution of wormholes surrounded by DZ profile halos}\label{ch: IV}
Let’s dive into the intriguing concept of wormholes and how we can describe them mathematically. A typical wormhole can be represented by the following metric:
\begin{equation}\label{cw1}
    ds^2=-e^{2\hat{\nu}(r)}dt^2+e^{2\hat{\mu}(r)}dr^2 +r^2(d\theta^2+\sin^2\theta d\phi^2).
\end{equation}
In this equation, $r$ is the radial coordinate, stretching from $r_0$ to infinity. The angular coordinates, $\theta$ and $\phi$, are limited to values between $0$ and $\pi$, and $0$ and $2\pi$, respectively. The function $\hat{\nu}$ is known as the redshift function, and it helps us understand how gravity influences light as it travels through these fascinating structures. One important feature of the function $\hat{\mu}(r)$ is its behavior at the wormhole's throat, where it tends to infinity:
\begin{eqnarray}
    \lim_{{r \to r_0^+}} \hat{\mu}(r) = +\infty.
\end{eqnarray}
Additionally, there's a crucial relationship between  $\hat{\mu}(r)$ and the shape function $\hat{\mathcal{S}}(r)$, which can be expressed as:
\begin{eqnarray}
    \hat{\mathcal{S}}(r)=r\left[1-e^{-2\hat{\mu}(r)}\right].
\end{eqnarray}
To fully understand a wormhole, we need to follow some important guidelines set by Morris and Thorne \citep{Morris:1988cz}. These guidelines focus on the shape function, denoted as $\hat{\mathcal{S}}(r)$.

First off, the metric coefficient $e^{2\hat{\nu}(r)}$ must be both finite and non-zero close to the throat of the wormhole, which we refer to as $r_0$. The shape function itself has to meet a couple of key conditions to ensure that the wormhole is traversable:

1. At the throat $r_0$, the shape function should equal the radial coordinate:
   \begin{eqnarray}
   \hat{\mathcal{S}}(r_0) = r_0.
   \end{eqnarray}

2. The rate at which the shape function changes at this point must not exceed 1:
    \begin{eqnarray}
   \hat{\mathcal{S}}'(r_0) \leq 1.
   \end{eqnarray}
   (Here, the prime symbol indicates we're talking about the derivative with respect to the radial coordinate.)

Additionally, there's an important inequality that the shape function must satisfy:
 \begin{eqnarray}
\hat{\mathcal{S}}'(r) < \frac{\hat{\mathcal{S}}(r)}{r}.
\end{eqnarray}
This condition ensures that the ratio of the shape function to the radial coordinate is decreasing, which helps avoid issues at the throat of the wormhole.

Lastly, when we look further away from the throat, we need to make sure that:
 \begin{eqnarray}
\hat{\mathcal{S}}(r) < r.
\end{eqnarray}

All these conditions work together to guarantee that the wormhole can be stable and safely traversed. They are crucial for any in-depth study of wormhole structures.

\begin{figure*}
\centering
\includegraphics[width=5.88cm,height=5.75cm]{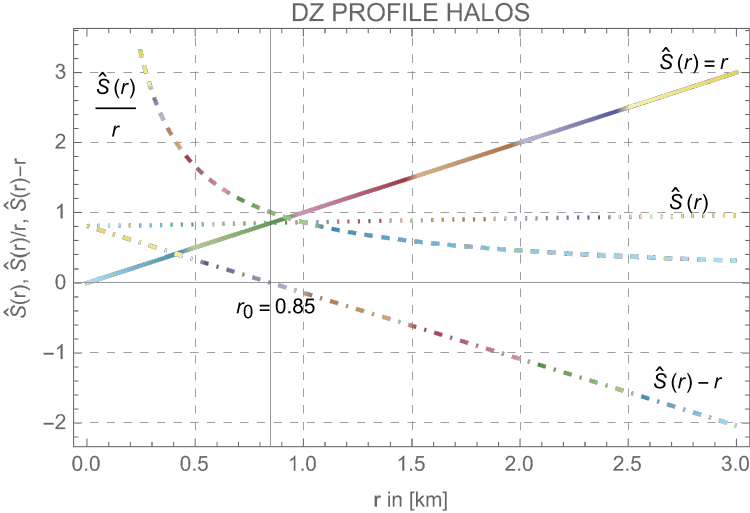}
\includegraphics[width=5.88cm,height=5.75cm]{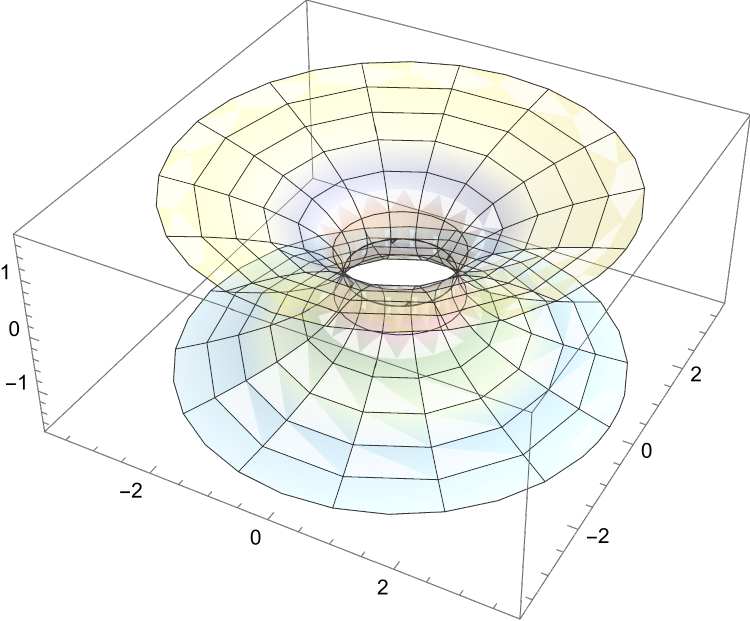}
\includegraphics[width=5.888cm,height=5.775cm]{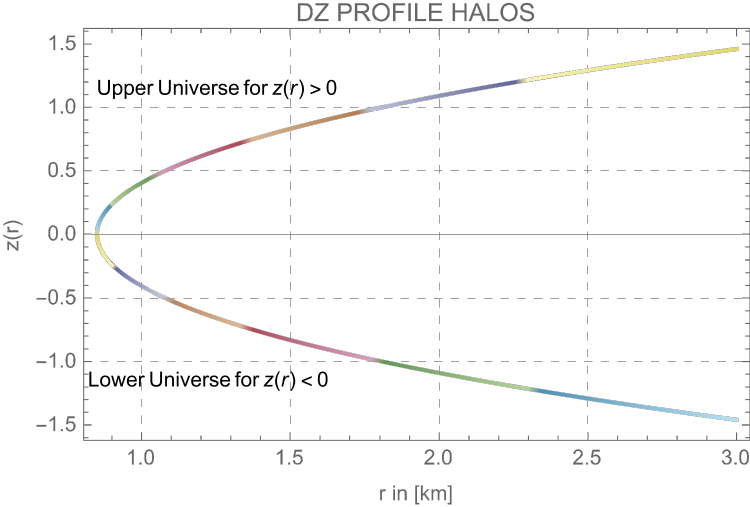}
\caption{The shape function for DZ profile halos features a consistent throat radius of $r_0 = 0.85$. The analysis considers several parameters, including $r_c = 0.5$, $\lambda = 0.2$, $\chi = 0.3$, $a = 0.2$, and $\rho_{ch} = 0.9$, highlighting how these factors influence the characteristics of the halos. }\label{fig1}
\end{figure*}

Next, we will delve into the non-null components of the field equations that govern the wormhole metric outlined in Eq. (\ref{cw1}) and the stress-energy tensor defined in Eq. (\ref{cw8}). Our focus will be on exploring these components within the context of $f(R, L_m, T)$ gravity, a framework that offers intriguing insights into the nature of these exotic structures.
\begin{eqnarray}\label{whfrlt1}
    -\frac{\hat{\mathcal{S}}}{r^3}&=&\left[\omega+\frac{\chi}{2}\right]P_{r}+\frac{\chi}{2}\left[\rho+2P_{t}\right], \\\label{whfrlt2}
\frac{1}{2r^2}\left[\frac{\hat{\mathcal{S}}}{r}-\hat{\mathcal{S}}'\right]&=&\left(\omega+\chi\right)P_{t}+\frac{\chi}{2}\left[\rho+P_{r}\right], \\ \label{whfrlt3}
    \frac{\hat{\mathcal{S}}'}{r^2}&=&\left[\omega +\frac{\chi}{2}\right]\rho+\frac{\chi}{2}\left[P_{r} +2P_{t}\right].   
\end{eqnarray}
In our analysis, we'll start by assuming that $\hat{\nu}(r)$ remains constant. To simplify our discussion, we define a new parameter, $\omega$, like this:
\begin{eqnarray}
\omega = 8\pi + \frac{\lambda}{2} + \chi.
\end{eqnarray}
With this in mind, we can take the earlier equations and rearrange them to clearly show what the matter content of the wormhole looks like.
\begin{eqnarray}\label{whfrlt4}
    \omega  r^2 \rho&=&\hat{\mathcal{S}}', \\\label{whfrlt5}
    \omega  r^3 P_{r}&=&-\hat{\mathcal{S}}, \\\label{whfrlt6}
    2 \omega  r^3 P_{t}&=&\hat{\mathcal{S}}-r \hat{\mathcal{S}}',  
\end{eqnarray}


\begin{figure}
\centering
\includegraphics[width=8.3cm,height=5.75cm]{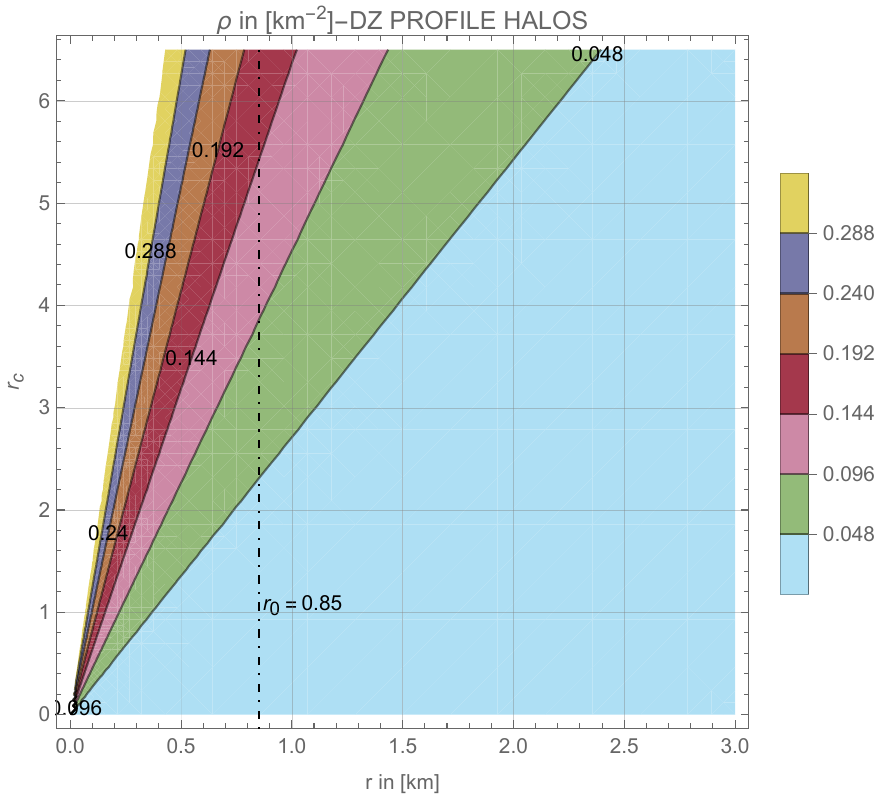}
\caption{ This diagram illustrates the matter density, $\rho$, for DZ profile halos, with the throat radius set at $r_0 = 0.85$. The plot feature specific parameter values, including $\lambda = 0.2$, $\chi = 0.3$, $a = 0.2$, and $\rho_{ch} = 0.9$, along with a range of values for $r_c$. }\label{fig2}
\end{figure}


\begin{figure}
\centering
\includegraphics[width=8.3cm,height=5.75cm]{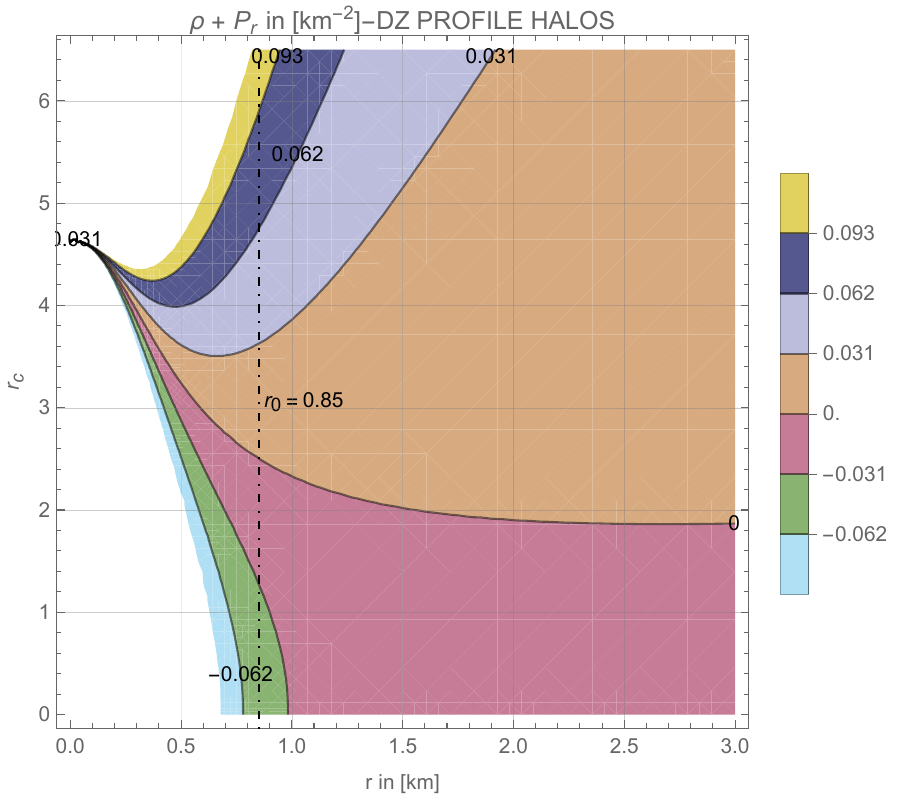}
\caption{This diagram illustrates the NEC, $\rho + P_r$, for DZ profile halos, with the throat radius set at $r_0 = 0.85$. The parameter values used in this plot are identical to those in Fig. \ref{fig2}.}\label{fig3}
\end{figure}

\begin{figure}[ht]
\centering
\includegraphics[width=8.3cm,height=5.75cm]{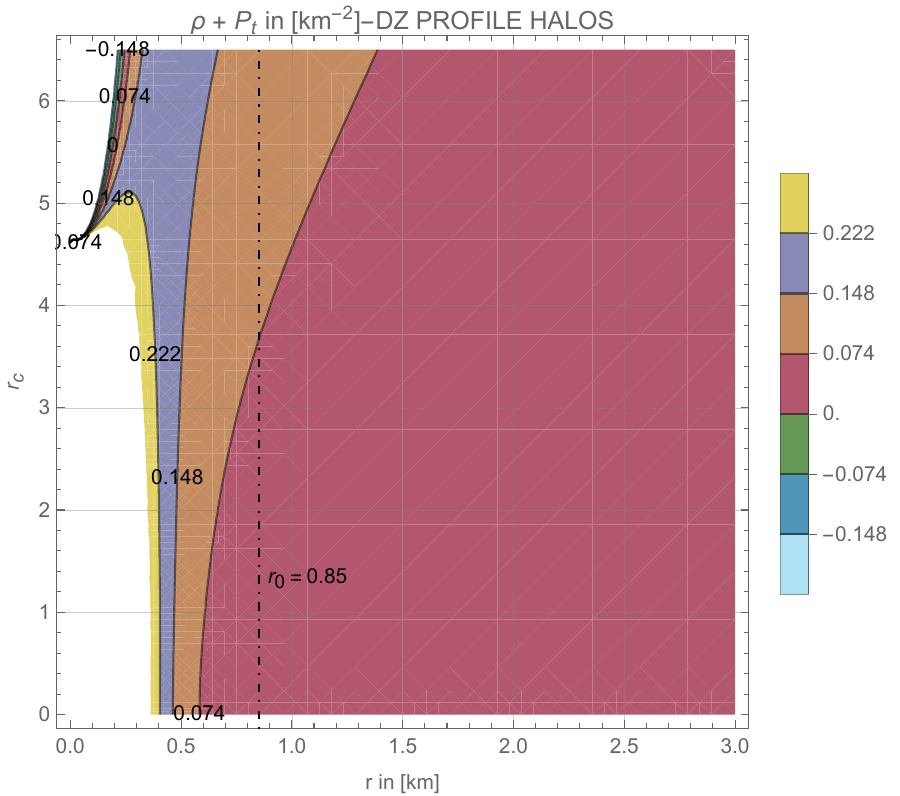}
\caption{ This diagram illustrates the NEC, $\rho + P_t$, for DZ profile halos, with the throat radius set at $r_0 = 0.85$. The parameter values used in this plot are identical to those in Fig. \ref{fig2}.}\label{fig4}
\end{figure}
\begin{figure}[ht]
\centering
\includegraphics[width=8.3cm,height=5.75cm]{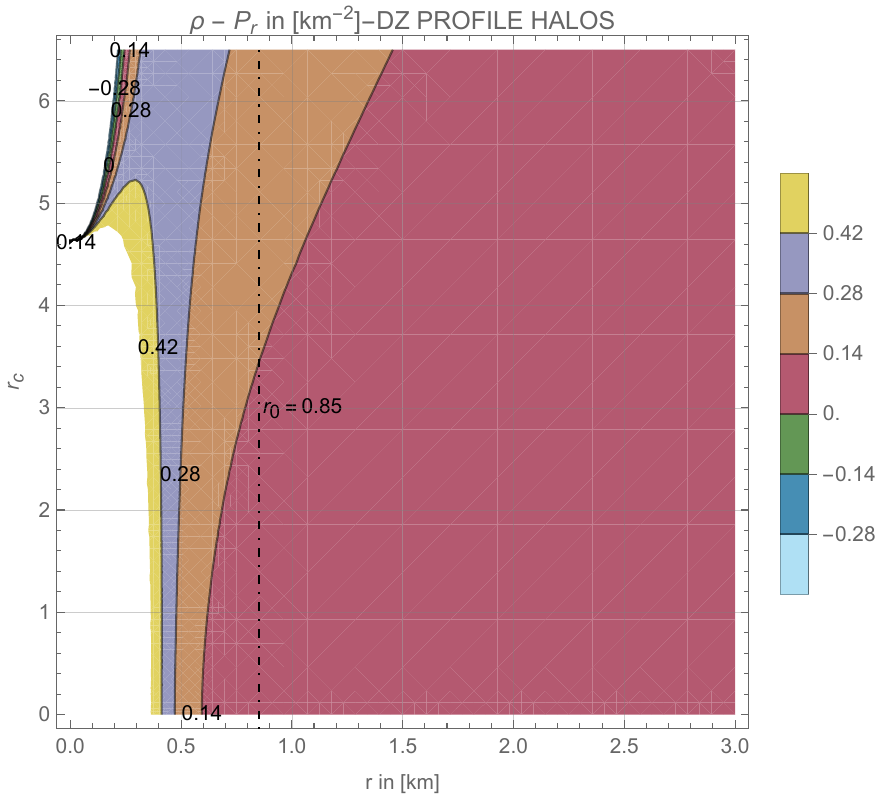}
\caption{ This diagram illustrates the DEC, $\rho - P_r$, for DZ profile halos, with the throat radius set at $r_0 = 0.85$. The parameter values used in this plot are identical to those in Fig. \ref{fig2}.}\label{fig5}
\end{figure}

In this paper, we explore the DZ profile, a flexible approach for modeling dark matter halos. This model helps researchers capture the diverse structural features found in astrophysical systems \cite{Zhao:1995cp,Zhao:1996mr}. The double power-law density profile smoothly shifts between different slopes at both small and large radii, making it well-suited for fitting observational data and aligning with theoretical predictions. More generally, Zhao \cite{Zhao:1995cp,Zhao:1996mr} shows that these double power-law density profiles can be described in the following way:
\begin{equation} \label{dmprofile1}
\rho(r) = \frac{\rho_{\rm ch}}{(r / r_{\rm c})^a \left(1 + (r / r_{\rm c})^{1/b}\right)^{b(g-a)}}.
\end{equation}
In our study of dark matter, we refer to $r_{\rm c}$ as the scale radius and $\rho_{\rm ch}$ as the characteristic density. The parameters $a$, $b$, and $g$ are essential as they shape the inner slope, how sharply the profile transitions, and the outer slope of the density curve. Interestingly, when we set $b = n$ and $g = 3 + k/n$ (with $n$ and $k$ being any natural numbers), we can derive straightforward formulas for important aspects like gravitational potential, enclosed mass, and velocity dispersion. The density profile is given by:
\begin{equation} \label{dmprofilev3}
  \rho(r) = \frac{\rho_{\rm ch}}{(r / r_{\rm c})^{a}\left(1 + (r / r_{\rm c})^{1 / 2}\right)^{2(3.5 - a)}}.
\end{equation}
By tweaking these parameters, the DZ profile can mimic a wide variety of observed dark matter distributions. This flexibility provides valuable insights into the nature and behavior of dark matter in our universe. We can find more details about the DZ profile in \cite{Freundlich:2020mr}.

In the following, we can derive the expression for the shape function by using Eqs. (\ref{whfrlt4}) to (\ref{whfrlt6}) along with Eq. (\ref{dmprofilev3}), which are specifically designed for zero-tidal-force wormholes. The resulting shape function is given by:
\begin{eqnarray}
 \hat{\mathcal{S}}(r) =\frac{2 \rho_{ch} r^3 \omega  \left(\sqrt{\frac{r}{r_{ch}}}+1\right)^{2a-6} \left(\frac{r}{_{ch}}\right)^{-a}}{6-2 a} + C.
\end{eqnarray}
To find the integrating constant $C$, we apply the condition $\hat{\mathcal{S}}(r_0) = r_0$. This leads us to the final form of the shape function:

\begin{small}
\begin{eqnarray}\label{sh}
 \hat{\mathcal{S}}(r) &=& \frac{1}{6-2 a}\Bigg[2 \rho_{ch} r^3 \omega  \left(\sqrt{\frac{r}{r_{ch}}}+1\right)^{2 a-6} \left(\frac{r}{r_{ch}}\right)^{-a}\nonumber\\&&-2 \rho_{ch} r_0^3 \omega  \left(\sqrt{\frac{r_0}{r_{ch}}}+1\right)^{2 a-6} \left(\frac{r_0}{r_{ch}}\right)^{-a}-2 (a-3) r_0\Bigg].~~~~~~~
\end{eqnarray}
\end{small}
Based on the shape function we derived in $(\ref{sh})$, we took a closer look at its features. This examination allowed us to create plots for $\hat{\mathcal{S}}(r)$, $\hat{\mathcal{S}}(r)/r$, and $\hat{\mathcal{S}}(r) - r$ specifically for DZ profile halos. We can see these plots illustrated in Fig. \ref{fig1}. By thoughtfully selecting the values for the remaining parameters--such as $r_0 = 0.85$, $r_c = 0.5$, $\lambda = 0.2$, $\chi = 0.3$, $a = 0.2$, and $\rho_{ch} = 0.9$--we can effectively produce the plots for the shape function. When we take a closer look at Fig. \ref{fig1}, we notice that for values of $r$ greater than $r_0$, the expression $\hat{\mathcal{S}}(r) - r$ becomes negative. This tells us that $\frac{\hat{\mathcal{S}}(r)}{r}$ is less than 1. Essentially, this means that as $r$ increases, $\hat{\mathcal{S}}(r) - r$ decreases, which meets the flaring-out condition for $r \geq r_0$ and ensures that $\hat{\mathcal{S}}'(r)$ remains below 1. Additionally, the criteria that $\hat{\mathcal{S}}'(r) < 1$ and $\frac{\hat{\mathcal{S}}(r)}{r} < 1$ hold true for all $r$ greater than $r_0$, along with the observation that $\frac{\hat{\mathcal{S}}(r)}{r}$ approaches 0 as $r$ goes to infinity, are consistently met across all distances. Thus, we can confidently say that spacetime shows asymptotic flatness at every radial distance. This leads us to conclude that the new shape function based on DZ profile halos effectively captures the essence of wormholes.

The embedding surface $Z(r)$ is defined by the following equation \cite{Morris:1988cz}:
\begin{equation}
      Z(r)=\pm \int_{\,r_0}^{\infty}\biggl(\frac{r}{\hat{\mathcal{S}}(r)}-1\biggr)^{-\frac{1}{2}}\text{d}r,
 \end{equation}
 where the two signs indicate the different branches of spacetime connected by the wormhole. This definition is important because it helps us grasp how the geometry of the wormhole changes as we navigate through spacetime. Continuing from our previous discussion on the flatness of spacetime, we've created visualizations of the wormhole configurations, which we can see in Fig. \ref{fig1}. This figure includes embedding diagrams that show the wormholes with DZ profile halos, generated by rotating them around the $z$-axis through a full $2\pi$ angle. These visualizations not only provide a clear picture of the wormholes but also illustrate the complex relationships between different regions of spacetime. In Fig. \ref{fig1}, we can observe the embedding diagram along with the corresponding embedded surface, representing the upper and lower universes in the context of DZ profile halos. The upper universe can be seen as the ``normal'' spacetime we experience, while the lower universe may represent alternate regions connected by the wormhole. These visuals are essential for deepening our understanding of the structure and topology of the wormholes, as they reveal how these fascinating structures might exist within a broader context of spacetime geometry.

\begin{figure}[ht]
\centering
\includegraphics[width=8.3cm,height=5.75cm]{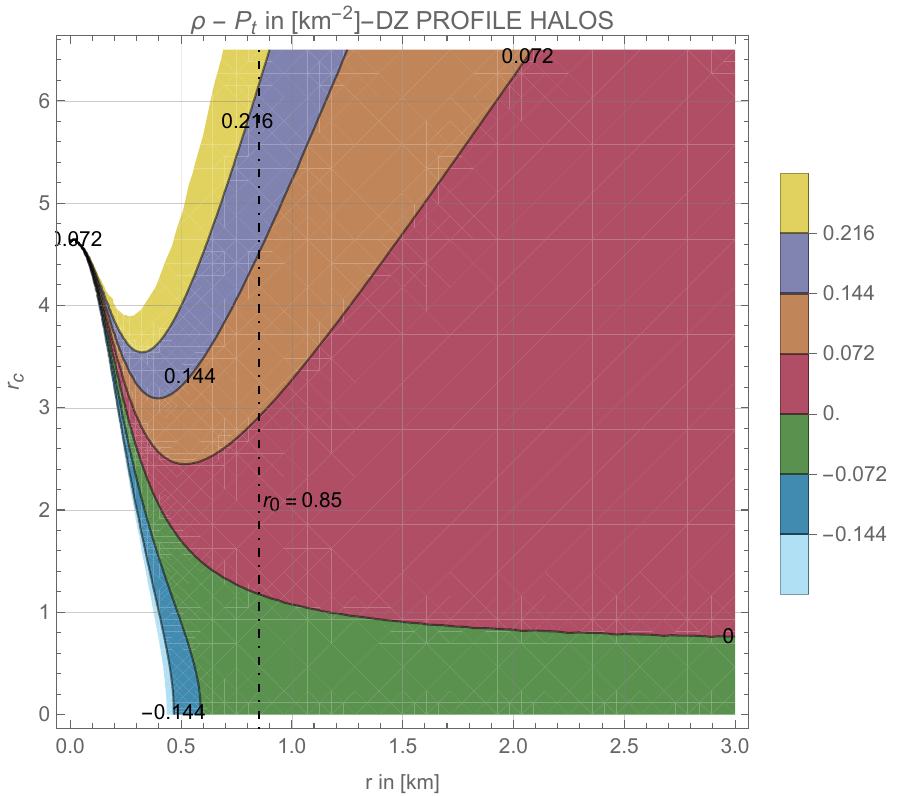}
\caption{ This diagram illustrates the DEC, $\rho - P_t$, for DZ profile halos, with the throat radius set at $r_0 = 0.85$. The parameter values used in this plot are identical to those in Fig. \ref{fig2}.}\label{fig6}
\end{figure}

In what follows, we can easily derive the expressions for the energy density and pressure components by plugging Eq. (\ref{sh}) into Eqs. (\ref{whfrlt4}) through (\ref{whfrlt6}). 
\begin{widetext}
\begin{small}
\begin{eqnarray}
    \rho(r)&=& \rho_{ch} \left(\sqrt{\frac{r}{r_{ch}}}+1\right)^{2 a-7} \left(\frac{r}{r_{ch}}\right)^{-a},\\
    P_r(r)&=&\frac{\rho_{ch} \left(\left(\sqrt{\frac{r}{r_{ch}}}+1\right)^{2 a-6} \left(\frac{r}{r_{ch}}\right)^{-a}-\frac{r_0^3 \left(\sqrt{\frac{r_0}{r_{ch}}}+1\right)^{2 a-6} \left(\frac{r_0}{r_{ch}}\right)^{-a}}{r^3}\right)}{a-3}-\frac{r_0}{r^3 \omega },\label{p_r}\\
    P_t(r)&=&-\frac{2 \rho_{ch} r^3 \omega  \left(\sqrt{\frac{r}{r_{ch}}}+1\right)^{2 a-7} \left(a+\sqrt{\frac{r}{r_{ch}}}-2\right) \left(\frac{r}{r_{ch}}\right)^{-a}-2 \rho_{ch} r_0^3 \omega  \left(\sqrt{\frac{r_0}{r_{ch}}}+1\right)^{2 a-6} \left(\frac{r_0}{r_{ch}}\right)^{-a}-2 (a-3) r_0}{4 (a-3) r^3 \omega }.\label{p_t}
\end{eqnarray}
\end{small}
\end{widetext}
To explore the ECs---specifically the NEC, WEC, DEC, and SEC---we've created graphs that highlight how each of these conditions behaves in relation to the energy density $\rho$, radial pressure $P_{r}$, and tangential pressure $P_{t}$. We can find these visual representations in Figs. \ref{fig2} to \ref{fig6}. In our analysis, we varied the model parameter $r_c$ and discovered that all the ECs can show both positive and negative values. This investigation sheds light on the important role of exotic matter in determining the throat radius of the wormhole geometry, offering valuable insights into the characteristics of the matter involved. For a more detailed understanding, we can dive deeper into the ECs using the formulas outlined in Eqs. (\ref{NEC})-(\ref{SEC}), as discussed in Sect. \ref{ch: III}.

In Fig. \ref{fig2}, we graphically represent the energy density $\rho$ as a function of the parameters $r_c$ and $r$. Our analysis shows that $\rho$ stays positive when $r_c$ is in the range of $[0, 6.5]$, given certain values of other free parameters. This is an important finding because it suggests that we can have physically reasonable matter distributions within the wormhole structure. In GR, having a positive energy density is crucial for the stability and feasibility of wormholes. It means that the matter supporting the wormhole does not lead to gravitational collapse and respects the necessary ECs. Thus, our results indicate that the configurations we've examined are not only theoretically interesting but could also represent stable and traversable wormholes in a way that makes physical sense.

Our model's findings, illustrated in Figs. \ref{fig3} to \ref{fig6}, provide important insights into the ECs at the throat of the wormhole. We found that the condition $\rho + P_{r}$ is completely violated when $r_c < 2.492$, but it holds true when $r_c > 2.492$. In contrast, the condition $\rho + P_{t}$ is consistently satisfied across the values we examined. This violation of $\rho + P_{r}$ at the throat for $r_c < 2.492$ indicates a breach of the NEC. This suggests the presence of exotic matter in this region, which is crucial for keeping the wormhole stable and preventing it from collapsing. We also observed that the DEC, represented by $\rho - P_{r}$, is fully satisfied at the throat in the radial direction (see Fig. \ref{fig5}). However, in the tangential direction ($\rho - P_{t}$), this condition is violated when $r_c < 1.201$ but holds true for $r_c > 1.201$ (see Fig. \ref{fig6}). This distinction shows how radial and tangential pressures behave differently in terms of the stability of the wormhole. Interestingly, the expression $\rho + P_{r} + 2P_{t} = 0$ tells us that the SEC is satisfied. This is significant because it helps us understand how matter and energy interact under the principles of GR. In summary, grasping these ECs is crucial for understanding how matter behaves in the universe and for advancing our knowledge about wormholes. Our findings not only enhance our theoretical understanding of wormholes but also highlight the importance of exotic matter in their potential existence and stability.

\section{The influence of gravitational lensing}\label{ch: V}
Let's start by looking at the Lagrangian for a null curve, which we can represent as $\gamma(s) = \bigl(t(s), r(s), \theta(s), \phi(s)\bigr)$. This curve exists within the spacetime framework described by Eq. (\ref{cw1}). The Lagrangian itself can be expressed as:
\begin{align}
    \mathcal{L}=-e^{2\hat{\nu}(r)}\frac{\Dot{t}^2}{2}+\left(1-\frac{\hat{\mathcal{S}}(r)}{r} \right)^{-1}\frac{\Dot{r}^2}{2}+r^2\frac{\Dot{\theta}^2}{2}+r^2\sin^2{\theta}\frac{\Dot{\phi}^2}{2}=0.
\end{align}
Given the spherical symmetry of the metric $ds^2$, we can simplify things a bit by setting $\theta = \frac{\pi}{2}$, which leads us to $\dot{\theta} = 0$. Now, let's define two important conserved quantities: energy $E$ and angular momentum $L$:
\begin{align}
  E:=\frac{\partial\mathcal{L}}{\partial \Dot{t}}=e^{-2\hat{\nu}}\Dot{t},~~\text{and}~~
  L:=\frac{\partial\mathcal{L}}{\partial \Dot{\phi}}=r^2\Dot{\phi}.
\end{align}
From these definitions, we can derive the effective potential:
\begin{align}
    E^2=V_{eff}(r)+e^{2\hat{\nu}}\left(1-\frac{\hat{\mathcal{S}}(r)}{r} \right)^{-1}\Dot{r}^2,
\end{align}
where 
\begin{align}
    V_{eff}(r):=e^{2\hat{\nu}}L^2/r^2.
\end{align}
\begin{figure}[ht]
\centering
\includegraphics[width=8.3cm,height=5.75cm]{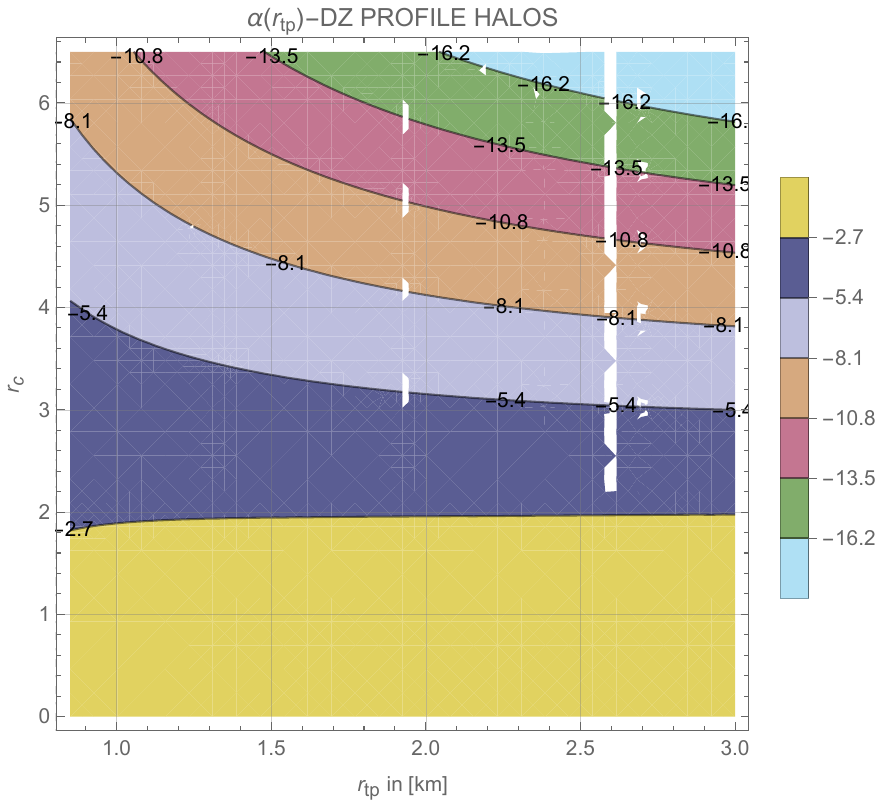}
\caption{ This diagram illustrates the deflection angle, $\alpha(r_{tp})$, for DZ profile halos, with the throat radius set at $r_0 = 0.85$. The parameter values used in this plot are identical to those in Fig. \ref{fig2}.}\label{fig9}
\end{figure}

By analyzing $V_{eff}$, we can explore the circular orbits associated with light-like paths, often called photon spheres. This is where things get interesting! When we take a closer look at the minima of $V_{eff}$, we notice that $\frac{dV_{eff}}{dr} \neq 0$ for a constant redshift. This suggests that there's a limitation related to our choice of coordinate system. To overcome this, we can switch to proper coordinates $l(r)$ at the non-singular point $r_0$. This adjustment helps us interpret the two distinct values of $l$ corresponding to a single $r$ as representing the two different spacetime regions connected by the wormhole.
\begin{equation}
l(r)=\pm\int_{\,r_0}^r\frac{1}{\sqrt{1-\frac{\hat{\mathcal{S}}(\epsilon)}{\epsilon}}}\text{d}\epsilon.
\label{eq: l}
\end{equation}
At $r = r_0$ when $l = 0$ \cite{V: SHA}, the potential hits its peak, suggesting that the throat behaves like an unstable photon sphere. To understand this better, let's take a look at the deflection angle $\alpha$ of the photon (see \cite{V: VIR, V: VIR1, V: VIR2}):
\begin{align}
    \alpha(r_{tp})=-\pi+2\int_{\,r_{tp}}^\infty\frac{e^{\hat{\nu}}}{\sqrt{1-\frac{\hat{\mathcal{S}}(r)}{r}}\sqrt{\frac{r^2}{\Psi^2}-e^{2{\hat{\nu}}}}}\text{d}r.
    \label{eq: DA}
\end{align}
In this equation, $\Psi := \frac{L}{E}$ represents the impact parameter. When we consider situations with a constant redshift, $\Psi$ has a straightforward linear relationship with the turning point $r_{tp}$, where the condition $\dot{r}(r_{tp}) = 0$ holds true:
\begin{equation}
\label{eq: u}
    \Psi=r_{tp}e^{-\hat{\nu}}.
\end{equation}
\begin{figure}[ht]
\centering
\includegraphics[width=8.3cm,height=5.75cm]{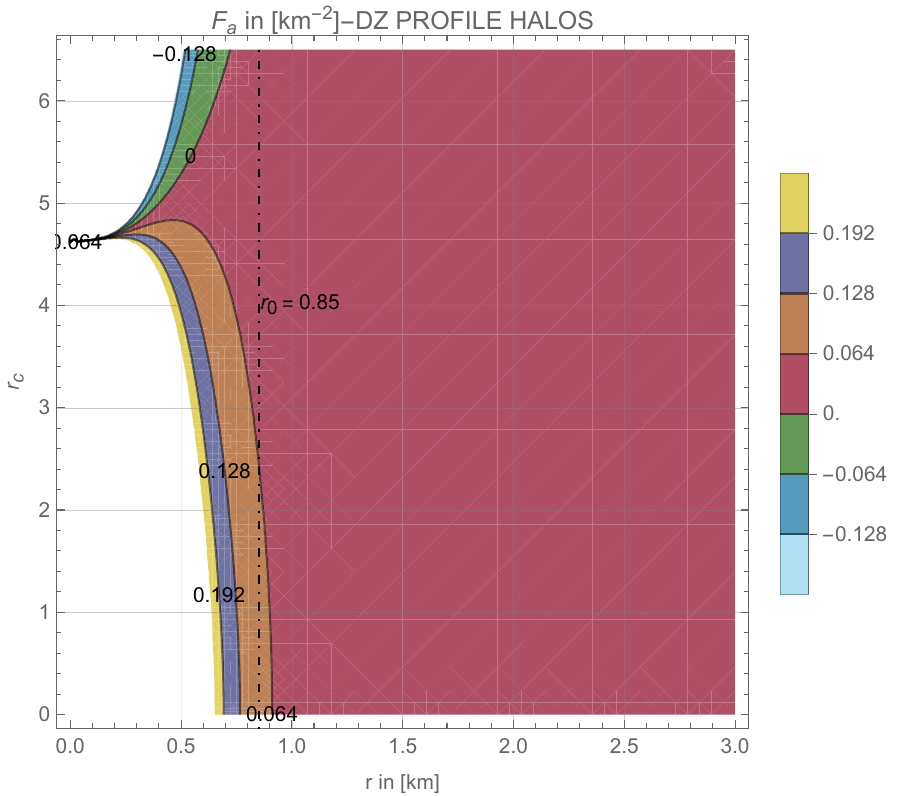}
\caption{ This diagram illustrates the anisotropic force, $F_a$, for DZ profile halos, with the throat radius set at $r_0 = 0.85$. The parameter values used in this plot are identical to those in Fig. \ref{fig2}.}\label{fig7}
\end{figure}
This means we can think of $\alpha$ as a function of the ratio $\frac{L}{E}$. Since $r = r_0$ is the only photon sphere in this system, the only singularity in the equation appears at $r_{tp} = r_0$. If $r_{tp}$ drops below $r_0$, the function starts to diverge. The deflection angle, called $\alpha(r_{tp})$, is shown in Fig. \ref{fig9} as a function of both $r_c$ and $r_{tp}$. This angle is strongly influenced by the shape of the wormhole and the value of $r_c$. When $r_c$ is between $0$ and $6.5$, the deflection angle stays negative for all corresponding values of $\alpha(r_{tp})$. This means that photons actually feel a repulsive gravitational force in this range. Instead of being pulled toward the wormhole, the light rays are pushed away, moving further from the throat of the wormhole. This behavior is intriguing because it shows a unique aspect of how gravity interacts with light around a wormhole. Rather than capturing light, the wormhole pushes it away, which changes our understanding of how light travels near these exotic structures. Such dynamics suggest that wormholes could significantly influence the way light behaves in their surroundings, challenging our traditional views of gravity and spacetime.

\section{Tolman-Oppenheimer-Volkoff equilibrium condition}\label{ch: VI}
The Tolman-Oppenheimer-Volkoff (TOV) equation, developed by Tolman, Oppenheimer, and Volkoff \cite{Oppenheimer}, plays a vital role in our understanding of gravitational equilibrium in spherically symmetric spacetimes. This equation is key to exploring the stability of different solutions in GR and modified theories of gravity \cite{Gorini}. Interestingly, it can also be adapted to consider anisotropic mass distributions \cite{Kuhfittig}, which broadens its relevance in the study of gravitational systems.
\begin{eqnarray}\label{51}
\frac{\nu'}{2}(\rho+p_r)+\frac{dp_r}{dr}+\frac{2}{r}(p_r-p_t)=0,
\end{eqnarray}

Based on Eq. \eqref{cw1}, we define $\nu = 2\hat{\nu}(r)$. This definition helps us clearly understand the roles of anisotropic, gravitational, and hydrostatic forces, which are expressed as:
\begin{equation}\label{52}
F_a=\frac{2}{r}(p_t-p_r), ~~F_g=-\frac{\nu'}{2}(\rho+p_r), ~F_h=-\frac{dp_r}{dr}.
\end{equation}

We achieve equilibrium when the equation $F_h+F_g+F_a=0$ holds. Interestingly, by choosing a constant redshift, this relationship simplifies even further, causing the gravitational force $F_g$ to effectively disappear:
\begin{equation}
\label{eq: eq}
    F_h+F_a=0.
\end{equation}
\begin{figure}[ht]
\centering
\includegraphics[width=8.3cm,height=5.75cm]{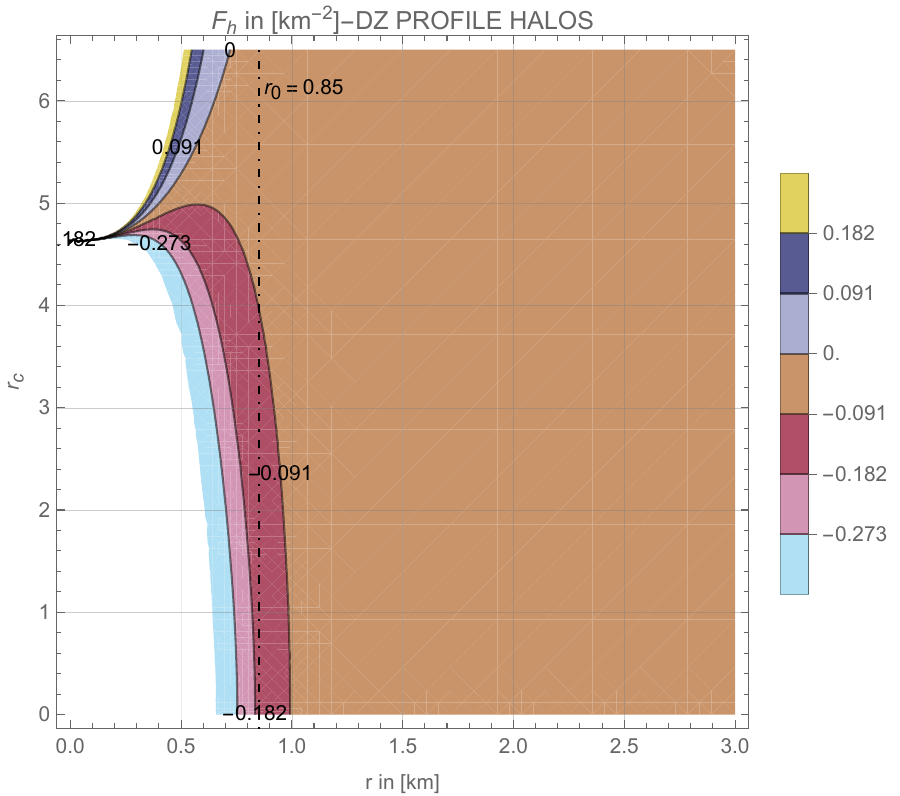}
\caption{ This diagram illustrates the hydrostatic force, $F_h$, for DZ profile halos, with the throat radius set at $r_0 = 0.85$. The parameter values used in this plot are identical to those in Fig. \ref{fig2}.}\label{fig8}
\end{figure}

In Figs. \ref{fig7} and \ref{fig8}, we illustrate the behavior of these forces for our DZ profile halos. Our findings show that the anisotropic force tends to be positive, while the hydrostatic force is negative, meaning they are acting against each other. For wormhole solutions to be in balance, it's essential that $F_{h} + F_{a} = 0$. This makes Eq. \eqref{eq: eq} an important criterion for assessing the stability of the solutions we've discussed. The fact that we observe equilibrium suggests that these forces are equal in strength but work in opposite directions. Therefore, we can confidently say that the wormhole solutions we identified for our DZ profile halos are stable.

\section{Conclusions}\label{ch: VII}
Studying traversable wormholes in GR opens up intriguing questions about exotic matter and its link to the NEC. In the late 1980s, Morris and Thorne made significant contributions by pointing out that exotic matter must meet the unusual condition of $p_r + \rho < 0$ for safe passage through these wormholes. In our research, we aim to go beyond traditional general relativity and delve into $f(R, L_m, T)$ modified gravity. This evolution allows us to explore how different gravitational theories might change the nature and feasibility of wormhole solutions. By deriving the gravitational field equations for this theory, we aim to uncover new ways to create traversable wormholes without relying on exotic matter. We adjust the matter Lagrangian and energy-momentum tensor using coupling strengths $\lambda$ and $\chi$, providing a method to reconcile matter and gravity. This flexibility offers exciting opportunities for understanding wormholes. Through our careful treatment of static wormhole solutions with a constant redshift function, we present our findings as a first-order approximation within this modified gravity framework. We based our wormhole shape function on the Dekel-Zhao dark matter distribution to ensure our solutions were traversable and compatible with the requirements for exotic matter, allowing wormholes to support both exotic and normal matter in this modified gravity scenario.

We found that our specific $f(R, L_m, T)$ function allows traversable wormholes to pass or fail energy condition tests in certain parameter spaces, with the matter stress-energy tensor sometimes meeting NEC requirements, demonstrating flexibility. To ensure the shape function behaves properly at infinity, we imposed restrictions on the coupling parameters. Our investigation of gravitational lensing revealed interesting effects, including a repulsive gravitational force for positive couplings, which influences the stability of our wormhole solutions and enhances our understanding of gravitational interactions in modified theories. Finally, using the TOV formalism, we confirmed the stability of our wormhole solutions, supporting their theoretical foundation and opening new research directions in gravitational physics and wormhole theory.

Our study paves the way for exciting future developments, such as exploring different forms of the $f(R, L_m, T)$ function to create wormhole solutions without exotic matter. Given the novelty of this theory, there is significant potential for innovation. One idea is to apply our findings to develop cosmological models, investigating how specific values of $\lambda$ and $\chi$ might describe an expanding universe. This could deepen our understanding of wormholes and greatly impact the broader discussion on modified gravity theories in cosmology.

\section*{Acknowledgements}
This research was funded by the Science Committee of the Ministry of Science and Higher Education of the Republic of Kazakhstan (Grant No. AP23487178). AE thanks the National Research Foundation of South Africa for the award of a postdoctoral fellowship.

\end{document}